\documentclass[aps,prl,showpacs,superscriptaddress,twocolumn,floatfix]{revtex4}
\usepackage{latexsym}
\usepackage{amsfonts}
\usepackage{amssymb}
\usepackage{amsmath}
\usepackage{graphicx}

\begin{document}

\title{Vortex oscillations induced by a spin-polarized current in a magnetic nanopillar: Evidence for a failure of the
Thiele approach}

\author{A. V. Khvalkovskiy}
\altaffiliation{Corresponding author. On leave from the A.M. Prokhorov General Physics Institute of RAS, Moscow, Russia. Electronic address: khvalkov@fpl.gpi.ru}
\affiliation{Unit\'e Mixte de Physique CNRS/Thales and Universit\'e
Paris Sud 11, RD 128, 91767 Palaiseau, France}
\affiliation{Istituto P.M. s.r.l., via Cernaia 24, 10122 Torino, Italy}
\author{J. Grollier}
\affiliation{Unit\'e Mixte de Physique CNRS/Thales and Universit\'e
Paris Sud 11, RD 128, 91767 Palaiseau, France}
\author{A. Dussaux}
\affiliation{Unit\'e Mixte de Physique CNRS/Thales and Universit\'e
Paris Sud 11, RD 128, 91767 Palaiseau, France} 
\author{K. A. Zvezdin}
\affiliation{Istituto P.M. s.r.l., via Cernaia 24, 10122 Torino, Italy}
\affiliation{A.M. Prokhorov General Physics Institute of RAS, Vavilova str. 38, 119991 Moscow, Russia}
\author{V. Cros}
\affiliation{Unit\'e Mixte de Physique CNRS/Thales and Universit\'e
Paris Sud 11, RD 128, 91767 Palaiseau, France}

\begin{abstract}
We investigate the vortex excitations induced by a spin-polarized current in a magnetic nanopillar by means of micromagnetic simulations and analytical calculations. Damped motion, stationary vortex rotation and the switching of the vortex core are successively observed for increasing values of the current. We demonstrate that even for small amplitude of the vortex motion, the analytical description based the classical Thiele approach can yield quantitatively and qualitatively unsound results. We suggest and validate a new analytical technique based on the calculation of the energy dissipation. 
\end{abstract}

\pacs{75.47.-m,75.40.Gb,85.75.-d}\maketitle

A magnetic vortex is a curling magnetization distribution, with the magnetization pointing perpendicular to the plane within the nanometer size vortex core. This unique magnetic object has attracted much attention recently because of the fundamental interest to specific properties of such a nanoscale spin structure. Gyrotropic modes of vortices in magnetic nanocylinders have been intensively studied theoretically \cite{GuslienkoJAP2002} and experimentally \cite{ParkCrowell}. Apart from their fundamental relevance, the unique properties of the vortices are of considerable practical interest for new applications to memory and microwave technologies. In this view, the reversal of the magnetization within the vortex core by magnetic field and spin-polarized current has been thoroughly studied \cite{VanWaeyenberge, Hertel, Yamada, GuslienkoPRL08}. More recently, sub-GHz dynamics of magnetic vortices induced by the spin transfer effect observed in nanopillars and nanocontacts \cite{Pribiag,Pufall,Mistral} have raised a strong interest. Indeed, the associated microwave emissions in such vortex-based Spin-Transfer Nano-Oscillators (STNOs) occur at low current densities, without external magnetic field, together with high powers and narrow linewidths ($<$ 1 MHz) comparatively to single-domain STOs. 

Traditionally, the analytical description of the vortex gyrotropic motion is based on the general approach for a translational motion of a magnetic soliton in an infinite media developed by A. Thiele \cite{Thiele}. This calculation consists in a convolution of the Landau-Lifshitz-Gilbert (LLG) equation with the magnetization distribution under a specific condition of a translational motion of the magnetization pattern. Eventually a single equation (often referred to as the Thiele equation) for the vortex core position {\bf X} can be derived. The approach developed by Thiele to build his equation has been used for a long time to derive equations of vortex motion in many magnetic systems. In particular, it is often used to describe analytically the vortex oscillations induced by spin current \cite{IvanovSTT,Liu,Kruger, Thiaville} in magnetic nanodiscs, in the '\textit{current perpendicular to the plane}'  (CPP) or '\textit{current in the plane}' (CIP) configurations. Vortex dynamics in magnetic submicron discs can not be considered as translational due to a strong deformation of the vortex structure by the edges \cite{GuslienkoMetlov}. Guslienko \textit{et al.} demonstrated that this deformation should be taken into account in the calculation of the system energy \cite{GuslienkoJAP2002}. However we show here that the impact of the spacial confinement on the vortex dynamics is much deeper. Our results evidence that, even if a proper model magnetization distribution is used, the Thiele approach applied for CPP nanodiscs with a spin current can give rise to significant qualitative and quantitative errors. We suggest a new analytical technique to estimate the spin current-induced effects in such systems. 

\begin{figure}[h]
   \centering
    \includegraphics[keepaspectratio=1,width=8.5 cm]{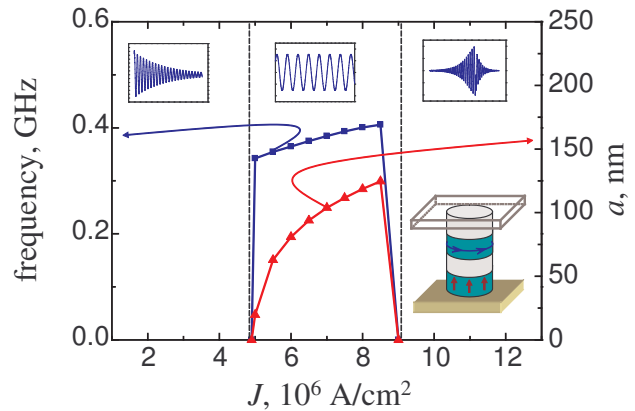}
     \caption{(color online) Steady vortex gyration induced by the spin polarized current: frequency $f$ = $ \omega / 2 \pi$ (squares) and radius of the vortex core orbit $a$ (triangles) as a function of the current density J (numerical simulations). Top insets: illustrations for the time evolution of the averaged projection of the free layer magnetization on the polar axis, for $J < J_{C1}$ (left), $J_{C1}< J < J_{C2}$ (center) and  $J > J_{C2}$ (right), in arbitrary units. Bottom inset: sketch of the device geometry.}
\label{Fig1}
\end{figure}

In our calculations we consider vortex motion in a vortex-STNO. The system under study is sketched in the inset of Fig. \ref{Fig1}. The nanopillar spin valve has a circular cross-section. The reference layer is a fixed perpendicular polarizer, the polarization vector $\mathbf{p}$ is perpendicular to the plane, and, to be clear in our interpretations, we disregard the stray magnetic field emitted by it. The initial magnetization distribution in the free layer is a vortex; the magnetization within the vortex core is parallel to $\mathbf{p}$. The current flow is assumed to be uniform in the pillar, with a radial symmetry of the current lines in the contact pads. The spin transfer term \cite{Slonc} in the calculations is given by $(\sigma J / M_s) \mathbf{M} \times (\mathbf{M} \times \mathbf{p})$, $J$ is the current density, $\mathbf{M}$ is the magnetization vector, $M_s$ is the magnetization of saturation and $\sigma$ represents the efficiency of the spin-transfer torque: $\sigma = \hbar P / (2 \left|e\right| L M_s)$, $P$ is the spin polarization of the current, e is the charge of the electron, $L$ is the sample thickness. 

As a starting point for our analytical calculations, we use a shortened form of the Thiele equation, which was used in Ref. \cite{GuslienkoJAP2002} to account for the frequency of low-amplitude vortex oscillations in magnetic nanodiscs:
\begin{equation}\label{ThieleEqTrunc}
\mathbf{G}\times\frac{d\mathbf{X}}{dt} - \frac{\partial W (\mathbf{X})}{\partial \mathbf{X}}=0
\end{equation}
Here the gyrovector $\mathbf{G}$ is given by:
\begin{equation}\label{GyroVector}
\mathbf{G} = - \frac{M_{s}}{\gamma} \int { \,d V \sin \theta \left( \bigtriangledown \theta \times \bigtriangledown \varphi \right)},
\end{equation}
$\gamma$ is the gyromagnetic ratio, $\theta, \varphi$ are the magnetization angles, and the integration in Eq. (\ref{GyroVector}) is over the magnetic disc. $W(\mathbf{X})$ is the potential energy of the shifted vortex. Guslienko \textit{et al.} \cite{GuslienkoJAP2002} showed that an appropriate model magnetization distribution for a moving vortex is given by  \textit{the two-vortices ansatz} (TVA):
\begin{eqnarray}\label{TwoVortexAnsatz}
\varphi\left(\rho,\chi;a,\chi_v\right) &=& g_a\left(\rho,\chi-\chi_v\right)+\chi_v, \nonumber \\
g_a\left(\rho,\chi\right) &=& tan^{-1}\left(\frac{\rho sin \chi}{\rho cos \chi -a}\right)  \\
&+& \tan^{-1}\left(\frac{\rho \sin \chi}{\rho \cos \chi -R^2 / a}\right)+C. \nonumber
\end{eqnarray}
Here $\rho$,$\chi$ are polar coordinates in the disc plane, $a$ = $\left|\textbf{X}\right|$ is the vortex core displacement, ($a$, $\chi_v$) is the position of the vortex core center, R is the radius of the dot and C = $\pi$/2 or C = $-\pi$/2 for different regions of the dot. The TVA Eq. (\ref{TwoVortexAnsatz}) defines a spin structure that satisfies the magnetostatic boundary conditions, i.e. assumes zero magnetic charges on the side borders of the disc. The out-of-plane magnetization component $M_s \cos \theta$ can be described by a bell-shaped function, that is non-zero in the core region few nanometers in diameter \cite{Usov}. Using Eq. (\ref{ThieleEqTrunc}), hereafter we address analytically the vortex gyrotropic motion with a frequency $f = \omega / 2 \pi $ and a small amplitude $a$:
\begin{equation}\label{GyroMotion}
\dot{\mathbf{X}} = 2\pi f \mathbf{e}_{z} \times \mathbf{X},\ a << R.
\end{equation}

It follows from Eq. (\ref{GyroVector}) that the gyrovector is given by $\mathbf{G}$ = $- G \mathbf{e_{z}}$, where $G$ = $2\pi M_s L / \gamma$ is the gyroconstant \cite{GuslienkoJAP2002}. At small current densities, the major contribution to the vortex energy W($\mathbf{X}$) is the magnetostatic energy $W_{m}$, arising from the appearance of volume magnetic charges for a shifted vortex. It is given by $W_{m}\left(a\right)$=$\frac{20}{9} \pi M_s^2 L^2 a^2 / R$ \cite{GuslienkoJAP2002}. Recent simulations have shown that the contribution of the Oersted magnetic field generated by the current can be very important \cite{YSChoi}. Therefore we also calculate the energy contribution $W_{Oe}$ due to the Oersted field. $W_{Oe}$ is given by integration of the energy density $\mathbf{-H_{Oe}(r) \cdot M(r,X)}$ over the volume, where $\mathbf{H_{Oe}(r)}$ is the Oersted field distribution at a given point $\mathbf{r}$. The integration yields $W_{Oe}\left(a\right)$ = $1.70 \pi L R M_s J a^2 / c $ \cite{OerstedNote}. Summing up these two contributions and using Eq. (\ref{ThieleEqTrunc}) and (\ref{GyroMotion}), one gets the analytical prediction for the vortex frequency:
\begin{equation}\label{Omega1}
f = f^{m}_{0}+f_{Oe}J,
\end{equation}
where $f^{m}_{0}= \frac{10}{9\pi}\gamma M_s L / R$ and $f_{Oe}=\frac{0.85}{\pi}\gamma R / c$. 

The critical current to launch the vortex oscillations can be found considering the energy balance in the system. The energy dissipation is given by $\dot{W}$ = $\int \left(\frac{\delta E}{\delta \theta}\dot{\theta} + \frac{\delta E}{\delta \varphi}\dot{\varphi}\right)$, where $\frac{\delta E}{\delta \theta}$ and $\frac{\delta E}{\delta \varphi}$ can be substituted by expressions found from the LLG equation as done in Ref. \cite{IvanovSTT}. Assuming that outside of the core region $\theta = \pi/2$ and neglecting the contribution from the core, one finds:
\begin{equation}\label{Dissipation2}
\dot{W} = - \alpha \frac{M_s}{\gamma} \int \dot{\varphi}^2 dV  + M_s  \int \left(\sigma J\right) \dot{\varphi}dV,
\end{equation}
$\alpha$ is the Gilbert damping. The first term in Eq. (\ref{Dissipation2}) corresponds to the energy loss due to the damping, and the second term to the dissipation arising from the spin transfer torque, which can be positive or negative according to the current sign. For a steady motion the energy is conserved ($\dot{W} = 0$), thus after some algebra, one can find from Eq. (\ref{Dissipation2}), (\ref{GyroMotion}):
\begin{equation}\label{Omega2}
4 \pi \alpha \eta f = \gamma \sigma J,
\end{equation}
where $\eta$ = $\frac{1}{2} ln(R/2c_e)-\frac{1}{8}$, $c_e$ is an integration limit, that is of the order of exchange length $l_e$ = $\sqrt{A/4 \pi M^2_s}$ \cite{Thiele}, $A$ is the exchange constant. From Eq. (\ref{Omega1}) and (\ref{Omega2}) we get an expression for the critical value $J_{C1}$ of the current density to excite the vortex oscillations:
\begin{equation}\label{JC1}
J_{C1}=\frac{\alpha \eta  f^{m}_{0}}{\gamma \sigma / 4 \pi - \alpha\eta f_{Oe}}.
\end{equation}

A different prediction follows from the Thiele approach. The Thiele equation that takes into account the damping and the spin transfer effect in the CPP configuration is given by \cite{IvanovSTT, Liu}:
\begin{equation}\label{ThieleEquation}
\mathbf{G}\times\frac{d\mathbf{X}}{dt} - \frac{\partial W}{\partial \mathbf{X}} - \hat{D} \frac{d\mathbf{X}}{dt} + \mathbf{F}_{ST}=0, 
\end{equation}
where the spin transfer force $\mathbf{F}_{ST}$ is:
\begin{equation}\label{FstThiele}
\mathbf{F}_{ST} = M_s L \int \left(\sigma J\right) \nabla \varphi \sin^2 \theta dV;
\end{equation}
(a different expression for $\mathbf{F}_{ST}$ has been derived for CIP systems \cite{Thiaville}). $\hat{D}$ is the damping tensor:
\begin{equation}\label{DampTensor}
\hat{D} = - \frac{M_{s} }{\gamma} \int { \,d V \left[ \bigtriangledown \theta \bigtriangledown \theta + \sin^{2} \theta \bigtriangledown \phi \bigtriangledown \phi \right]}.
\end{equation}
For circular dots, $\hat{D}$ = $D \hat{E} $, where $\hat{E}$ is a unit tensor and the damping constant is $D$ = $\alpha \eta' G$ \cite{GuslienkoD}. The factor $\eta'$ and the previously introduced $\eta$ define the same quantity even if they are given by different expressions as we discuss below. Calculation of the spin transfer force for the TVA yields $\mathbf{F}_{ST}$ = $2 \pi M_s L \sigma J a \mathbf{e_{\chi}}$ \cite{IvanovSTT, Liu}. For a steady gyrotropic motion, the third and the last terms of Eq. (\ref{ThieleEquation}) are perpendicular to the first and the second terms. Therefore the frequency of the vortex motion, given by Eq. (\ref{ThieleEqTrunc}), is not affected by the new terms of Eq. (\ref{ThieleEquation}); instead, they define the amplitude of the vortex motion \cite{IvanovSTT}. For the steady motion, the damping term is balanced by $\mathbf{F}_{ST}$, thus one gets: 
\begin{equation}\label{Omega2Thiele}
2 \pi \alpha \eta' f = \gamma \sigma J.
\end{equation}
Comparing this to Eq. (\ref{Omega2}), we see that Eq. (\ref{Omega2Thiele}) yields about a twice smaller value of the first critical current $J_{C1}$. The reason of this difference is related to the breakdown of the assumption of a translational motion for the vortex, thus of the basic underlying assumption for the Thiele approach. 

Indeed, the derivation of Eq. (\ref{GyroVector}), (\ref{DampTensor}) and (\ref{FstThiele}) essentially uses the following feature of the translational motion of a magnetic soliton: $\dot{\mathbf{M}}$ = $-\left(\dot{\mathbf{X}}, \nabla \right) \mathbf{M}$ \cite{IvanovSTT}. For a vortex moving in a nanodisc, however, the left-hand side of this expression vanishes at the disc side border, but the right-hand side has a finite value. It can be shown that for $a$ $<<$ $R$, most contribution to the magnitude of the gyroforce $\mathbf{G}$ and the damping $\hat{D}$ comes from the vicinity of the disc center ($\rho < R /2$); the magnetization motion within his region can be considered translational, to some approximation. In contrast to it, even for a very small $a$, about a half of the magnitude of $\mathbf{F}_{ST}$ given by Eq. (\ref{FstThiele}) originates from the region of the disc close to its boundary.

Our analytical results can be compared to numerical micromagnetic simulations. In the simulations a nanopillar 300 nm in diameter is considered. The free layer is 10 nm thick and has the following magnetic parameters: $M_s$ = 800 emu/cm$^3$, A = 1.3 $\times$ 10$^{-6}$ erg/cm and $\alpha$ = 0.01 (values for permalloy). We use a two-dimensional mesh with in-plane cell size 1.5 $\times$ 1.5 nm$^2$. The polarization is taken to be P = 0.2. The micromagnetic simulations are performed by numerical integration of the LLG equation using our micromagnetic code based on the forth order Runge-Kutta method with an adaptive time-step control for the time integration. 

We observe vortex excitations only for positive current, which is defined as a flow of electrons from the free layer to the polarizer. The vortex motion is damped for small current densities $J \leq J_{C1}$, where the first critical current density $J_{C1}$ $=$ 4.9 $\times$ 10$^6$ A/cm$^2$. For larger currents, after some transitional period, the vortex is gyrating on a steady circular orbit. Interestingly, $J_{C1}$ is about one order of magnitude less than the critical current density for excitation of magnetization oscillations in nanopillar STNOs with nominally uniform free magnetic layer \cite{Kiselev, IC1note}. The values of $J$, for which the steady vortex oscillations are observed, are limited by the second critical current value $J_{C2}$ $= $9.0 $\times$ 10$^6$ A/cm$^2$. For J $\geq J_{C2}$, on reaching a critical orbit, the core of the vortex is reversed. The details of this process: appearance of a vortex with opposite polarity and an antivortex, annihilation of the latter with the original vortex, essentially reproduce the previous findings for the vortex core switching by the field or current \cite{Liu,Hertel,Yamada}. After the reversal of the core, the direction of the vortex gyration is changed and the vortex oscillations are damped. 

For each point within $J_{C1} < J < J_{C2}$, the vortex motion is simulated for 100 ns after reaching a stationary orbit. The vortex frequencies extracted from these simulations together with the radius of the oscillation orbit are presented in Fig.\ref{Fig1}. On increasing $J$, the oscillation frequency increases ranging from 0.34 GHz to 0.41 GHz. The radius of the orbit increases with the current as well, reaching 125 nm at $J$ $=$ 8.5 $\times$ 10$^6$ A/cm$^2$. 

Analytical prediction for the vortex frequency at $J$ = $J_{C1}$, given by Eq. (\ref{Omega1}), is f = 0.36 GHz that is in a good correspondance to the simulation results $f$ = 0.34 GHz. This is in agreement with the previous results of Ref. \cite{GuslienkoJAP2002}, which showed that Eq. (\ref{ThieleEqTrunc}) can properly describe the frequency of small-amplitude vortex oscillations in magnetic nanodots. The factor $\eta$ for our system can be extracted using additional simulations. We find that if the current is switched off, the vortex motion is a gyration with the orbit damped in time as $a$ $\propto$ $\exp\left(- t / \tau \right)$, $\tau$ is a time constant. On the other hand, it follows from Eq. (\ref{Dissipation2}) and (\ref{GyroMotion}) that at zero current $a$ $\propto$ $exp\left(-2 \alpha \pi \eta f t\right)$ \cite{EtaNote}. Comparing this to the numerical results, we find $\eta$ = $\eta'$ = 1.65. This value is in reasonable agreement with the prediction of the current work ($\eta$ = 1.34 taking $c_e$ = $l_e$) and previously made analytical estimations (e.g. $\eta'$ = 2.3 from Ref. \cite{GuslienkoD}). 

Taking $\eta$ and $f$ from the simulations, we find from Eq. (\ref{Omega2}) that the analytical prediction $J_{C1}$ $=$ 4.5 $\times$ 10$^6$ A/cm$^2$ is in good agreement with the numerical result. The result of the Thiele approach (Eq. (\ref{Omega2Thiele})) is $J_{C1}$ $=$ 2.3 $\times$ 10$^6$ A/cm$^2$, that illustrates our statement of its failure to account for the vortex motion in CPP nanopillars. 

\begin{figure}[h]
   \centering
    \includegraphics[keepaspectratio=1,width=8.5 cm]{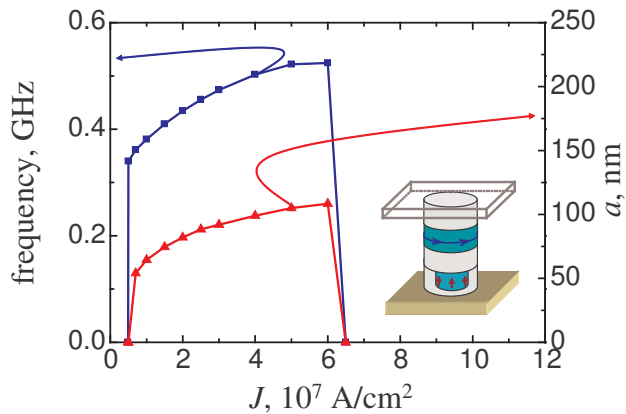}
     \caption{ (color online) Numerical result for the constrained polarizer, in the notations of Fig. \ref{Fig1}. Inset: sketch of the device geometry.}
\label{Fig2}
\end{figure}

We demonstrate the breakdown of the Thiele approach in a different way. We perform simulations for a configuration that we call constrained polarizer, for which the current polarization P equals to 0.2 for $\rho$ $\leq$ 50 nm and P = 0, hence $\sigma = 0$, for larger $\rho$. Thus in contrast to the case of the uniform polarizer for this new structure, the current does not excite the regions close to the disc boundary. The resulted dependences $f$($J$) and $a$($J$) are presented in Fig. \ref{Fig2}. We find for this configuration that $J_{C1}$ = (4.9 $\pm$ 0.1) $\times $10$^6$ A/cm$^2$ equals to the critical current for a uniform polarizer. This is in perfect agreement with the analytical result from the energy consideration. Indeed, Eq. (\ref{Dissipation2}) gives equal results, hence equal values of $J_{C1}$, for both configurations at $a$ $<<$ $R$. 

The prediction of the Thiele approach for a constrained polarizer is that the magnitude of the spin-transfer force $\mathbf{F}_{ST}$ (Eq. (\ref{FstThiele})) is a factor 1.8 less than that for a uniform polarizer. Accordingly, the value of $J_{C1}$ in the two configurations differs at about the same factor, in contradiction to the numerical results. Thus we demonstrate that the Thiele approach can give rise to not only quantitative, but also qualitative disagreements.

Simulation results for the constrained polarizer contain other remarkable facts. We see that the second critical current $J_{C2}$ = 6.5 $\times$10$^7$ A/cm$^2$ is by a factor of about 7 larger that $J_{C2}$ for a uniform polarizer.  The oscillation orbit gradually increases with the current reaching $a$ = 108 nm at $J$ = $J_{C2}$. The oscillation frequency starts at f = 0.34 GHz at $J_{C1}$ like for a uniform polarizer but reaches a larger frequency f = 0.52 GHz at $J$ = $J_{C2}$. 

It has been predicted that the vortex core is reversed if its velocity reaches a critical value $v_{crit}$ that is about 340 m/s for permalloy independently of the device design \cite{GuslienkoPRL08}. Our simulation results are 320 m/s for the uniform polarizer and 350 m/s for the constrained polarizer in nice agreement with this prediction. A small difference between the values is presumably due to a different extent of the vortex deformation at critical orbits. The value of $v_{crit}$ together with the dependencies of $f(J)$ and $a(J)$ define the value of the second critical current $J_{C2}$. As for the constrained polarizer $f(J)$ and $a(J)$ are much less steep functions than those for the uniform polarizer, $J_{C2}$ in the former is substantially larger than in the latter. Larger frequencies at a given orbit, found for the constrained polarizer, are related to the strong influence of the Oersted field. These facts make this configuration promising for applications. It can be implemented by reducing the polarizer dimensions or simply by using a point contact technique. 

In conclusion, we clearly show that the Thiele formalism fails to give a sound analytical description for the vortex motion in the CPP magnetic nanopillars due to the breakdown of the underlying assumption of the translational motion of the vortex. We suggest a new original analytical approach based on calculation of the energy dissipation, which is shown to be in good agreement with the numerical results. Our calculations demonstrate that vortex-based STNOs can potentially have very low values of $J_{C1}$, large values of $J_{C2}$ and operate at zero external magnetic field, which make them promising candidates for future microwave technology applications. 

The work is supported by the EU project MASTER (grant 212257) and RFBR (grants 09-02-01423 and 08-02-90495).


\begin{thebibliography}{32}
\bibitem{OerstedNote}
The details of this calculation and verification will be given in a forthcoming publication. We note that for $J \leq 10^{7}$ A/cm$^{2}$ the contribution of W$_{Oe}$ to the total density is relatively small (1\%-10\%), however it can become predominant for larger J.
\bibitem{EtaNote}
From Eq. (\ref{ThieleEquation}) it follows $a$ $\propto$ $exp\left(-2 \alpha \pi \eta' f t\right)$ at $J$ = 0, thus $\eta$ and $\eta'$ are equivalent quantities.
\end{thebibliography}
\end{document}